\documentstyle[11pt,amssymb
]{article}

\newcommand{\Tr}{\mathop{\rm Tr}\nolimits}

\title{Couplings of heavy hadrons with soft pions\\ from QCD sum rules}

\author{A.~G.~Grozin$^{1,2}$ and O.~I.~Yakovlev$^{1,3}$\\
Budker Institute of Nuclear Physics,\\
Novosibirsk 630090, Russia}

\date{}

\begin{document}

\maketitle

\begin{abstract}
We estimate the couplings in the Heavy Hadron Chiral Theory (HHCT) lagrangian
from the QCD sum rules in an external axial field.
Stability of the sum rules at moderate values of the Borel parameter
is poor that probably signals a slow convergence of the OPE series.
At large values of the Borel parameter they stabilize, and yield
the couplings much lower than the constituent quark model expectations.
This region is not trustworthy for baryons,
but in the meson case only unexpectedly large contributions
of a few lowest excitations could invalidate the prediction $g_1\approx0.2$.
\end{abstract}

\addtocounter{footnote}{1}
\footnotetext{Supported in part by grant from the Soros foundation,
and by grant from the Soros foundation
awarded by the American Physical Society}
\addtocounter{footnote}{1}
\footnotetext{Internet address: \tt GROZIN@INP.NSK.SU}
\addtocounter{footnote}{1}
\footnotetext{Internet address: \tt O\_YAKOVLEV@INP.NSK.SU}

\section{Introduction}

It is well known that the QCD lagrangian with $n_l$ massless flavours has
the $SU(n_l)_L\times SU(n_l)_R$ symmetry spontaneously broken to $SU(n_l)_V$
giving the $(n_l^2-1)$-plet of pseudoscalar massless Goldstone mesons (pions)
$\pi^i_j$ ($\pi^i_i=0$). Their interactions at low momenta are described
by the chiral lagrangian (see e.~g.~\cite{Georgi})
\begin{equation}
L_\pi = \frac{f^2}{8}\Tr \partial_\mu \Sigma^+ \partial_\mu \Sigma + \cdots,
\quad
\Sigma = \exp \frac{2i\pi}{f},
\label{Lpi}
\end{equation}
where the pion constant $f\approx132$MeV is defined by
\[
{<}0|j^i_{j\mu}|\pi{>} = if e^i_j p_\mu, \quad
j^i_{j\mu} = \overline{q}_j \gamma_\mu\gamma_5 q^i \to
i\frac{f^2}{2} \left(\Sigma^+\partial_\mu \Sigma
- \Sigma\partial_\mu \Sigma^+\right)^i_j
\]
($e^i_j$ is the pion flavour wave function), and dots mean terms with more
derivatives. Light quark masses can be included perturbatively, and lead
to extra terms in~(\ref{Lpi}). $SU(n_l)_L\times SU(n_l)_R$ transformations
act as $\Sigma \to L\Sigma R^+$. Let's define $\xi=\exp i\pi/f$,
$\Sigma=\xi^2$; it transforms as $\xi \to L\xi U^+ = U\xi R^+$ where $U$
is a $SU(n_l)$ matrix depending on $\pi(x)$. The vector
$v_\mu=\frac12(\xi^+\partial_\mu \xi+\xi\partial_\mu \xi^+)$ and the axial
vector $a_\mu=\frac{i}{2}(\xi^+\partial_\mu \xi-\xi\partial_\mu \xi^+)$
transform as $v_\mu \to U(v_\mu+\partial_\mu)U^+$, $a_\mu \to U a_\mu U^+$.
There is a freedom in transformation laws of matter fields such as $\psi^i$
because it is always possible to multiply them by a matrix depending on $\pi$.
The only requirement is the correct transformation with respect to $SU(n_l)_V$
($L=R$). It is convenient to choose $\psi\to U\psi$. Then the covariant
derivative $D_\mu=\partial_\mu+v_\mu$ transforms as $D_\mu\psi\to UD_\mu\psi$.
Covariant derivatives of tensors with more flavour indices are defined
similarly.

Hadrons with a heavy quark are now successively investigated in the framework
of the Heavy Quark Effective Theory (HQET)~\cite{HQET} (for review and
references see~\cite{HQETrev}). To the leading order in $1/m$, the heavy
quark spin does not interact and can be rotated or switched off at all
(spin-flavour and superflavour symmetry). The $\overline{Q}q$ mesons with a
spinless heavy quark form the $j^P=\frac12^+$ $n_l$-plet $\psi^i$. The $Qqq$
baryons can have $j^P=0^+$ or $1^+$ giving the scalar flavour-antisymmetric
$n_l(n_l-1)/2$-plet $\Lambda^{ik}$ and the vector flavour-symmetric
$n_l(n_l+1)/2$-plet $\vec{\Sigma}\,^{ik}$. Switching the heavy quark spin on
gives degenerate $0^-$ and $1^-$ $B$ and $B^*$ mesons,
$\frac12^+$ $\Lambda$ baryons, and degenerate $\frac12^+$ and $\frac32^+$
$\Sigma$ and $\Sigma^*$ baryons.

Interaction of these ground-state heavy hadrons with soft pions is described
by the Heavy Hadron Chiral Theory~\cite{HHCT,Cho}. Excited mesons were
incorporated in~\cite{ExMes}, and electromagnetic interactions---in~\cite{Rad};
chiral loop effects were considered in~\cite{Loop,Cho}. We start from the HHCT
lagrangian with the heavy quark spin switched off:
\begin{eqnarray}
&&L = L_\pi + \overline{\psi}_i iD_0 \psi^i
  + \Lambda^*_{ij} iD_0 \Lambda^{ij}
  + \vec{\Sigma}\,^*_{ij} \cdot (iD_0-\Delta) \vec{\Sigma}\,^{ij}
\label{LHHCT}\\
&& + g_1 \overline{\psi}_i \hat{a}\,^i_j \gamma_5 \psi^j
  + 2ig_2 \vec{\Sigma}\,^*_{ik} \cdot \vec{a}\,^i_j \times \vec{\Sigma}\,^{jk}
  + 2g_3 \left( \Lambda^*_{ik} \vec{a}\,^i_j \cdot \vec{\Sigma}\,^{jk}
        + \vec{\Sigma}\,^*_{ik} \cdot \vec{a}\,^i_j \Lambda^{jk} \right),
\nonumber
\end{eqnarray}
where $\Delta$ is the $\Sigma$--$\Lambda$ mass difference. The possibility
of consideration of the $\Sigma\Lambda\pi$ interaction in HHCT relies on
the fact that this difference is small compared to the chiral symmetry
breaking scale though formally both of them are of the order of the
characteristic hadron mass scale. The matrix elements of the axial current
between heavy hadrons are easily obtained using PCAC:
\begin{eqnarray}
&&{<}M'|\vec{\jmath}\,^i_j|M{>} = g_1 \overline{u}'_j \vec{\gamma}\gamma_5 u^i,
\label{PCAC}\\
&&{<}\Sigma'|\vec{\jmath}\,^i_j|\Sigma{>}
 = 2ig_2 \vec{e}\,^{\prime*}_{jk} \times \vec{e}\,^{ik}, \quad
{<}\Lambda|\vec{\jmath}\,^i_j|\Sigma{>}
 = 2g_3 e^*_{jk} \vec{e}\,^{ik},
\nonumber
\end{eqnarray}
where $u^i$, $e^{ij}$, $\vec{e}\,^{ij}$ are the $M$, $\Lambda$, $\Sigma$
wave functions, and the nonrelativistic normalization of the states and
wave functions is assumed. If we switch the heavy quark spin on, we obtain
the usual HHCT lagrangian~\cite{HHCT,Cho}.

The HHCT couplings $g_i$ should be in principle calculable in the underlying
theory---HQET, but this is a difficult nonperturbative problem.
Some experimental information is available only on $g_1$. If we neglect
$1/m_c$ corrections, then
$\displaystyle \Gamma(D^{*+}\to D^0\pi^+) = \frac{g_1^2 p_\pi^3}{6\pi f^2}$,
and similarly for $D^+\pi^0$ (with the extra $1/2$). The experimental
upper limit~\cite{ACCMOR} on $\Gamma(D^{*+})$ combined with the branching
ratios~\cite{CLEO} give $g_1<0.68$. A combined analysis of $D^*$ pionic and
radiative decays was performed in~\cite{Rad}; it gives $g_1\sim0.4$--$0.7$.

In the constituent quark model, $g_1$ is the axial charge $g$ of the
constituent light quark in the heavy meson. Moreover, following the
folklore definition ``constituent quark is $B$ meson minus $b$ quark'',
this is the most clear way to define $g$ of the constituent quark.
The baryonic couplings $g_{2,3}$ are also equal to $g$ in this model.
The most naive estimate is $g\approx1$; the nucleon axial charge is
$g_A=\frac53g$ in the constituent model, and in order to obtain $g_A=1.25$
we should assume $g=0.75$.

Sum rules~\cite{SVZ} were successfully used to solve many nonperturbative
problems in QCD and HQET. The currents with the quantum numbers of the ground
state mesons and baryons with the heavy quark spin switched off are
\begin{equation}
j^i_M=Q^*\frac{1+\gamma_0}{2}q^i, \quad
j^{ij}_{\Lambda1}=(q^{T[i}C\gamma_5 q^{j]})Q, \quad
\vec{\jmath}\,^{ij}_{\Sigma1}=(q^{T(i}C\vec{\gamma}q^{j)})Q,
\label{Curr}
\end{equation}
where $Q$ is the spinless static quark field, $C$ is the charge conjugation
matrix, $q^T$ means $q$ transposed, $(ij)$ and $[ij]$ mean symmetrization
and antisymmetrization. There are also currents $j^{ij}_{\Lambda2}$
and $\vec{\jmath}\,^{ij}_{\Sigma2}$ with the additional $\gamma_0$.
Correlator of the mesonic currents was investigated in~\cite{Shuryak,BrGr},
and of the baryonic ones---in~\cite{GrYak}. The sum rules results are in a
qualitative agreement with the constituent quark model: the massless quark
propagator plus the quark condensate contribution simulate the constituent
quark propagation well enough.

The sum rules method was generalized to the case of a constant external field
for calculation of such static characteristics of hadrons as the magnetic
moments~\cite{IoSm}. Sum rules in an external axial field were used%
{}~\cite{BelKog} for calculation of $g_A$ of light baryons. In the present
work we use HQET sum rules in an axial field to calculate $g_{1,2,3}$.
Sum rules for the $D^*D\pi$ coupling have been studied earlier~\cite{ElKog};
their result (in the terms used here) is $g_1=0.2$.
We disagree with some results of this paper.

\section{Mesons}

We introduce the external axial field $A^i_{j\mu}$ ($A^i_{i\mu}=0$)
by adding the term
\begin{equation}
\Delta L=j^i_{j\mu}A^j_{i\mu}
\label{Field}
\end{equation}
to the lagrangian. We are going to calculate correlators of the currents%
{}~(\ref{Curr}) up to the terms linear in $A$ (these terms are denoted by
the subscript $A$). The light quark propagator in the gauge $x_\mu A_\mu(x)=0$
gets the contribution
\begin{equation}
S^i_{jA}(x,0)=-\frac{iA^i_j\cdot x}{2\pi^2}
\left(\frac{\hat{x}\gamma_5}{x^4}
-\frac{x_\mu\tilde{G}_{\mu\nu}\gamma_\nu}{4x^2}\right)+\cdots
\label{Prop}
\end{equation}
The $G^2$ term in $S_A$ vanishes after the vacuum averaging; we are not
going to calculate gluonic contributions beyond $G^2$ and hence may omit
this term.\footnote{The first term in~(\ref{Prop}) was presented
in~\cite{BelKog}, but we don't understand how the authors calculated
$G^2$ corrections without the second term and the above statement
about the third one.}
The axial field induces the quark condensates\footnote{These formulae
were presented in~\cite{BelKog} but with a different second term in the
first equation; their term does not satisfy the relation
${<}(\hat{D}q)\overline{q}{>}=0$ that follows from the equation
of motion.}
\begin{eqnarray}
{<}q^{ia\alpha}(x)\overline{q}_{jb\beta}(0){>}_A &=& \frac{\delta^a_b}{4N}
\Big\{\Big[ f^2\hat{A}\,^i_j
+ \frac{i}{6}{<}\overline{q}q{>}
\left([\hat{A}\,^i_j,\hat{x}] + 6 A^i_j\cdot x \right)
\nonumber\\
&& - \frac{m_1^2 f^2}{36} \left( 5 x^2 \hat{A}\,^i_j
- 2 A^i_j\cdot x\hat{x} \right) \Big] \gamma_5 \Big\}^\alpha_\beta,
\label{Qcond}\\
{<}q^{ia\alpha}g\tilde{G}\,^A_{\mu\nu}\overline{q}_{jb\beta}{>}_A &=&
\frac{\left(t^A\right)^a_b}{12C_{\rm F}N} m_1^2 f^2
\left(A^i_{j\mu}\gamma_\nu-A^i_{j\nu}\gamma_\mu \right)^\alpha_\beta,
\nonumber
\end{eqnarray}
where $m_1^2\approx0.2\mbox{\rm GeV}^2$ is defined by~\cite{Use}%
\footnote{the sum rule considered in the second paper of~\cite{Use}
yields the relation $m_1^2\approx m_0^2/4$.}
${<}0|\overline{d}g\tilde{G}_{\mu\nu}\gamma_\nu\gamma_5u|\pi^+{>}
=im_1^2fp_\mu$, $N=3$ is the number of colours, $C_{\rm F}=\frac{N^2-1}{2N}$.
We assumed $p\cdot A=0$ where $p\to0$ is the momentum of the field $A$;
in general these formulae should contain $A_\bot=A-\frac{A\cdot p}{p^2}p$.

The correlator of the meson currents~(\ref{Curr}) has the $A$-term
\begin{equation}
i{<}T j^i(x) \overline{\jmath}_j(0){>}_A = \frac{1+\gamma_0}{2} \hat{A}\,^i_j
\gamma_5 \frac{1+\gamma_0}{2} \delta(\vec{x}) \Pi_A(x_0)
\label{Mcor}
\end{equation}
that depends only on $\vec{A}\,^i_j$ and not $A^i_{j0}$.
The correlator possesses the usual dispersion representation at any
$A^i_{j\mu}$. The meson contribution at $A^i_{j\mu}=0$ is
$\rho(\omega)=F^2\delta(\omega-\varepsilon)$ where $\varepsilon$ is
the meson energy and ${<}0|j^i|M{>}=iFu^i$ (the usual meson constant is
$f_M=\frac{2F}{\sqrt{m}}$). Switching on the external field produces
the energy shift
$\varepsilon\to\varepsilon-g_1\gamma_5\vec\gamma\cdot\vec{A}$
($s_n=\frac12\gamma_5\vec\gamma\cdot\vec{n}=\pm\frac12$ is the meson spin
projection). Therefore the meson contribution is
$\rho_A(\omega)=g_1F^2\delta'(\omega-\varepsilon)+c\delta(\omega-\varepsilon)$
where the second term originates from the change of $F^2$. Besides that
there is a smooth continuum contribution $\rho_A^{\rm cont}(\omega)$.
Thus we obtain
\begin{equation}
\Pi_A(\omega)=\frac{g_1 F^2}{(\omega-\varepsilon)^2}
+\frac{c}{\omega-\varepsilon}+\Pi_A^{\rm cont}(\omega).
\label{Phys}
\end{equation}
In other words, the lowest meson's contribution in both channels
(Fig.~\ref{Fphys}a) gives the double pole at $\omega=\varepsilon$;
mixed lowest--higher state contributions (Fig.~\ref{Fphys}b) give a single
pole at $\omega=\varepsilon$ plus a term with a spectral density in the
continuum region after the partial fraction decomposition; higher states'
contributions (Fig.~\ref{Fphys}c) have a spectral density in the continuum
region only.

We can also calculate the correlator using OPE. Gluons don't interact with
the heavy quark in the fixed point gauge $x_\mu A_\mu(x)=0$. The diagrams
with non-cut light quark line (Fig.~\ref{Fmes}a, b) vanish. The diagram
with cut line (Fig.~\ref{Fmes}c) gives
\begin{equation}
\Pi_A(t) = \frac{f^2}{4} \left( 1 - \frac{i{<}\overline{q}q{>}t}{3f^2}
- \frac{5}{36}m_1^2 t^2 \right).
\label{OPEmes}
\end{equation}
Thus the appearance of $g_A$ of the constituent quark is entirely caused by
interaction with the quark condensate.
The correlator~(\ref{OPEmes}) corresponds to the spectral density in the form
of $\delta(\omega)$ and its derivatives.
Equating these expressions at an imaginary time $t=-i/E$ we obtain the sum rule
\begin{equation}
\left(\frac{g_1F^2}{E}+c\right)e^{-\varepsilon/E} = \frac{f^2}{4}
\left(1-\frac{{<}\overline{q}q{>}}{3f^2E}+\frac{5}{36}\frac{m_1^2}{E^2}\right).
\label{SRmes}
\end{equation}
At sufficiently high energies the hadronic spectral density is equal to
the theoretical one, and it vanishes. Moreover, higher
states' contributions are exponentially suppressed, therefore we neglect
them. This sum rule should agree with the $m_c\to\infty$ limit of the sum rule
for the $D^*\to D\pi$ coupling~\cite{ElKog}. We disagree with the last
term in~\cite{ElKog}, though it is not very important numerically.

At large $E$ the sum rule~(\ref{SRmes}) leads to
\begin{equation}
g_1 F^2 = {\textstyle\frac14}f^2\varepsilon
- {\textstyle\frac1{12}}{<}\overline{q}q{>}.
\label{Dualmes}
\end{equation}
Of course, higher states' contributions are not exponentially suppressed
at large $E$. But the contribution of high energy excitations is given
by the perturbative spectral density which is zero in our case. Hence
only one or few lowest states contribute. Excited mesons are supposed
to have $F^2$ substantially smaller than the ground state meson, and may be
are not very important. So we may hope that~(\ref{Dualmes}) gives
us a reasonable estimate. We use $F^2=(240\mbox{\rm MeV})^3$ and
$\varepsilon=430\mbox{\rm MeV}$ obtained from
the sum rule without radiative corrections~\cite{Shuryak} because we
don't know radiative corrections to~(\ref{OPEmes}) and feel that it
would be inconsistent to take them into account in a part of sum rules
(this would lead to large errors e.~g.\ in the sum rule for the Isgur-Wise
function). Of course, the existence of large radiative corrections~\cite{BrGr}
leads to an additional uncertainty in the result. If higher states'
contributions to~(\ref{Dualmes}) are not important, we obtain $g_1\approx0.2$.

We can multiply~(\ref{SRmes}) by $\exp(\varepsilon/E)$ and differentiate
in $E$ in order to exclude $c$. The result is shown in Fig.~\ref{Fgmes}.
Of course, at large enough $E\gtrsim2$GeV~(\ref{Dualmes}) is reproduced,
and we get $g_1\approx0.2$. This is the only region in which the sum rule
is stable. A similar sum rule with the finite $m_c$ was analyzed
in~\cite{ElKog} in the region $E=1.5$--3GeV with the same result.
But this requires an extremely slow convergence of the OPE in order to
forbid using lower $E$. The OPE series~(\ref{SRmes}) looks like
$1+200\mbox{\rm MeV}/E+(170\mbox{\rm MeV}/E)^2$, so it seems that
the expansion is applicable at $E>500$MeV. This is close to
the lower bound of the applicability region of the ordinary sum rule%
{}~\cite{Shuryak}. The upper bound of its applicability region is about
800MeV because the continuum contribution is too large at higher $E$.
If we suppose that the applicability region of the sum rule~(\ref{SRmes})
is approximately the same, then we obtain $g_1\sim0.4$--$0.7$. Stability of
the sum rule in this region is poor. In this case we have to assume
that the higher states' contribution is significant at $E>800$MeV.

\section{Baryons}

Correlators of the $\Sigma\Sigma$ and $\Lambda\Sigma$ currents have the
$A$-terms
\begin{eqnarray}
i{<}T j^{ij}_{\Sigma l}(x) j^*_{\Sigma i'j'm}(0){>}_A &=&
i\varepsilon_{lmn}A^{(i}_{n(i'}\delta^{j)}_{j')}\delta(\vec{x})\Pi_A(x_0),
\nonumber\\
i{<}T j^{ij}_\Lambda(x) \vec{\jmath}\,^*_{\Sigma i'j'}(0){>}_A &=&
\vec{A}\,^{[i}_{(i'}\delta^{j]}_{j')}\delta(\vec{x})\Pi_A(x_0).
\label{Bcor}
\end{eqnarray}
If we define ${<}0|j^{ij}_\Lambda|\Lambda{>}=F_\Lambda e^{ij}_\Lambda$,
${<}0|\vec{\jmath}\,^{ij}_\Sigma|\Sigma{>}=F_\Sigma\vec{e}\,^{ij}_\Sigma$,
then the physical states' contributions to the correlators (Fig.~\ref{Fphys})
are
\begin{eqnarray}
\Pi_A(\omega)&=&\frac{2g_2F_\Sigma^2}{(\omega-\varepsilon_\Sigma)^2}
+\frac{c}{\omega-\varepsilon_\Sigma}+\cdots
\label{Bphys}\\
\Pi_A(\omega)&=&\frac{2g_3F_\Lambda F_\Sigma}
{(\omega-\varepsilon_\Lambda)(\omega-\varepsilon_\Sigma)}
+\frac{c_\Lambda}{\omega-\varepsilon_\Lambda}
+\frac{c_\Sigma}{\omega-\varepsilon_\Sigma}+\cdots
\nonumber
\end{eqnarray}
It is impossible to separate the $g_3$ term from the mixed
$\Lambda$-excited and $\Sigma$-excited contributions unambiguously.
We can do it approximately if
$\Delta=\varepsilon_\Sigma-\varepsilon_\Lambda\ll
\varepsilon_c-\varepsilon_{\Lambda,\Sigma}$
because in such a case partial fraction decomposition of the first term would
give large contributions to $c_{\Lambda,\Sigma}$ $\sim1/\Delta$
with the opposite signs while the natural scale of $c_{\Lambda,\Sigma}$
in~(\ref{Bphys}) is
$c_{\Lambda,\Sigma}\sim 1/(\varepsilon_c-\varepsilon_{\Lambda,\Sigma})$.
This is not a defect of the sum rule but the uncertainty inherent to $g_3$
which can be defined only when $\Delta$ is small compared to
the chiral symmetry breaking scale.
We choose to require $c_\Lambda=c_\Sigma$; the choices $c_\Lambda=0$
or $c_\Sigma=0$ would be equally good.

The baryonic correlators in the OPE framework are described by the diagrams
in Fig.~\ref{Fbar}. The diagrams Fig.~\ref{Fbar}a--c with the non-cut quark
line interacting with the axial field vanish due to~(\ref{Prop}). We use
the factorization approximation for the four-quark condensate in Fig.%
{}~\ref{Fbar}e. In this approximation two diagonal correlators
${<}j_1j_1{>}$ and ${<}j_2j_2{>}$ coincide in both $\Sigma\Sigma$ and
$\Lambda\Sigma$ cases, as well as ${<}j_{\Lambda1}j_{\Sigma2}{>}$ and
${<}j_{\Lambda2}j_{\Sigma1}{>}$. This is similar to the usual heavy baryon
sum rules~\cite{GrYak}, and confirms the observation that
$F_{\Lambda1}=F_{\Lambda2}$ and $F_{\Sigma1}=F_{\Sigma2}$ within the
factorization approximation to the sum rules. Only even-dimensional
condensates contribute to the diagonal $\Sigma\Sigma$ and the nondiagonal
$\Lambda\Sigma$ correlators:
\begin{equation}
\Pi_A(t)=\frac{2N!\,f^2}{N\pi^2t^3}\left[1
-\left(\frac{5}{3}\pm\frac{C_{\rm B}}{C_{\rm F}}\right)\frac{m_1^2t^2}{12}
+\frac{\pi^2{<}\overline{q}q{>}^2t^4}{6Nf^2}\right],
\label{OPEbar1}
\end{equation}
where $C_{\rm B}/C_{\rm F}=1/(N-1)$ (this term comes from the diagram
Fig.~\ref{Fbar}f). Only odd-dimensional condensates contribute to the
nondiagonal $\Sigma\Sigma$ and the diagonal $\Lambda\Sigma$ correlators:
\begin{equation}
\Pi_A(t)=\frac{2N!\,{<}\overline{q}q{>}}{3N\pi^2t^2}\left[1-
\frac{3\pi^2f^2t^2}{2N}
\left(1+{\textstyle\frac1{16}}m_0^2t^2-{\textstyle\frac5{36}}m_1^2t^2\right)
\right].
\label{OPEbar2}
\end{equation}
These correlators correspond to the spectral densities
\begin{equation}
\rho_A(\omega)=\frac{N!\,f^2}{N\pi^2}\left[\omega^2+
\left(\frac53\pm\frac{C_{\rm B}}{C_{\rm F}}\right)\frac{m_1^2}{6}
\right],
\quad
\rho_A(\omega)=-\frac{2N!\,{<}\overline{q}q{>}\omega}{3N\pi^2}
\label{rho}
\end{equation}
(plus $\delta(\omega)$ and its derivatives).

We use the standard continuum model
$\rho_A^{\rm cont}(\omega)=\rho_A^{\rm theor}\vartheta(\omega-\varepsilon_c)$.
Equating the OPE~(\ref{OPEbar1}, \ref{OPEbar2}) and the spectral
representation at $t=-i/E$, we obtain the sum rules
\begin{eqnarray}
&&\left(\frac{2g_2F_\Sigma^2}{E}+c\right)e^{-\varepsilon_\Sigma/E}
\nonumber\\
&&\quad=\frac{4f^2}{\pi^2}E^3\left[f_2\left(\varepsilon_c/E\right)
+\frac{13m_1^2}{72E^2}f_0\left(\varepsilon_c/E\right)
+\frac{\pi^2{<}\overline{q}q{>}^2}{18f^2E^4}\right]
\nonumber\\
&&\quad=-\frac{4{<}\overline{q}q{>}}{\pi^2}E^2
\left[f_1\left(\varepsilon_c/E\right)
+\frac{\pi^2f^2}{2E^2}
\left(1-\frac{m_0^2}{16E^2}+\frac{5m_1^2}{36E^2}\right)
\right],
\nonumber\\
&&\left(\frac{2g_3F_\Lambda F_\Sigma}{\Delta}\mathop{\rm th}\frac{\Delta}{2E}
+\frac{c}{2}\right)
\left(e^{-\varepsilon_\Lambda/E}+e^{-\varepsilon_\Sigma/E}\right)
\label{SRbar}\\
&&\quad=\frac{4f^2}{\pi^2}E^3\left[f_2\left(\varepsilon_c/E\right)
+\frac{7m_1^2}{72E^2}f_0\left(\varepsilon_c/E\right)
+\frac{\pi^2{<}\overline{q}q{>}^2}{18f^2E^4}\right]
\nonumber\\
&&\quad=-\frac{4{<}\overline{q}q{>}}{\pi^2}E^2
\left[f_1\left(\varepsilon_c/E\right)
+\frac{\pi^2f^2}{2E^2}
\left(1-\frac{m_0^2}{16E^2}+\frac{5m_1^2}{36E^2}\right)
\right],
\nonumber
\end{eqnarray}
where $f_n(x)=1-e^{-x}\sum_{m=0}^{n}\frac{x^m}{m!}$.

Fig.~\ref{Fgbar}a shows the results for $g_{2,3}$ obtained from
the diagonal $\Sigma\Sigma$ and the nondiagonal $\Lambda\Sigma$ correlators
(the first formulae in~(\ref{SRbar})).
The values of $\varepsilon_{\Lambda,\Sigma}$ and $F_{\Lambda,\Sigma}$
are taken from the diagonal sum rules~\cite{GrYak}
in the middle of the stability plato $E\approx300$MeV.
We use the same values of the continuum threshold as in~\cite{GrYak}:
$(5.6\pm0.5)k$ for $\Sigma$ and $(4.6\pm0.5)k$ for $\Lambda$
where $k^3=-\frac{\pi^2}{6}{<}\overline{q}q{>}$, $k=260$MeV.
We put the continuum threshold in the sum rule for $g_2$
($\Sigma\Sigma$ correlator) equal to that in the $\Sigma$ sum rule;
in the case of $g_3$ ($\Lambda\Sigma$ correlator)
we put it equal to the average of the thresholds
in the $\Lambda$ and $\Sigma$ channels.
Of course, the effective continuum threshold for the $A$-term
in the correlators does not necessary coincide with that
in the absence of the external field;
this assumption gives an additional uncertainty.
The spectral density~(\ref{rho}) behaves like $\omega^2$,
and hence the continuum contribution to the sum rules~(\ref{SRbar})
quickly grows: it is equal to the result at $E\approx600$MeV
and is three times larger than the result at $E\approx900$MeV.
If we assume that the uncertainty in the continuum contribution
is, e.~g., 30\%, then we can't trust the sum rules at $E>900$MeV.
The lower bound of the applicability region is determined by
the convergence of the OPE series for the correlators.
It behaves like $1+(190\mbox{\rm MeV}/E)^2+(245\mbox{\rm MeV}/E)^4$
for the $\Sigma\Sigma$ correlator;
in the $\Lambda\Sigma$ case $140$MeV enters instead of $190$MeV.
It seems that OPE should be applicable at $E>400$MeV.
Stability of the sum rules in this window is poor.
We can only guess that $g_2\sim0.2$--1, $g_3\sim0.1$--1,
and probably $g_3<g_2$.

Fig.~\ref{Fgbar}b shows the results for $g_{2,3}$ obtained from
the nondiagonal $\Sigma\Sigma$ and the diagonal $\Lambda\Sigma$ correlators
(the second formulae in~(\ref{SRbar})).
The values of $\varepsilon_{\Lambda,\Sigma}$ and $F_{\Lambda,\Sigma}$
are taken from the nondiagonal sum rules~\cite{GrYak}
in the middle of the stability plato $E\approx650$MeV.
Again we use the continuum thresholds from~\cite{GrYak}:
$(5.6\pm0.5)k$ for $\Sigma$ and $(3.65\pm0.5)k$ for $\Lambda$.
The spectral density~(\ref{rho}) behaves like $\omega$,
and the continuum contribution grows not so quickly:
it is equal to the result at $E\approx1$GeV
and is three times larger than the result at $E\approx1.6$GeV.
The OPE series behaves like
$1+(290\mbox{\rm MeV}/E)^2\left[1-(150\mbox{\rm MeV}/E)^2\right]$,
and the applicability region starts at a larger $E$.
Stability of these sum rules is poor too.
They tend to prefer somewhat larger values $g_2\sim g_3\sim0.5$--1.

In conclusion, poor stability of the sum rules at moderate values
of the Borel parameter probably signals a slow convergence of the OPE series.
At large values of the Borel parameter they stabilize,
and yield the couplings $g_{1,2,3}$ much lower
than the constituent quark model expectations.
This region is not trustworthy for baryons
due to large continuum contributions.
In the meson case, higher excited states are dual to the vanishing
theoretical spectral density, and don't contribute.
Only unexpectedly large contributions of a few lowest excitations
could invalidate the prediction $g_1\approx0.2$.
This inconclusive picture is especially surprising
because similar sum rules in an external axial field
produce $g_A=1.25$ for the nucleon~\cite{BelKog}
in excellent agreement with experiment and the constituent quark model.

{\bf Acknowledgements}. We are grateful to V.~L.~Chernyak for useful
discussions.



\begin{figure}[p]
\vspace{5mm}
\caption{Physical states' contributions to the correlator}
\label{Fphys}
\end{figure}

\begin{figure}[p]
\caption{Correlator of meson currents}
\label{Fmes}
\end{figure}

\begin{figure}[p]
\caption{Sum rule for $g_1$}
\label{Fgmes}
\end{figure}

\begin{figure}[p]
\caption{Correlator of baryon currents}
\label{Fbar}
\end{figure}

\begin{figure}[p]
\caption{Sum rules for $g_{2,3}$ (solid and dashed lines):
a---diagonal $\Sigma\Sigma$ and nondiagonal $\Lambda\Sigma$ sum rules;
b---nondiagonal $\Sigma\Sigma$ and diagonal $\Lambda\Sigma$ sum rules.}
\label{Fgbar}
\end{figure}


\begin{thebibliography}{99}
\raggedright
\newcommand{\PLB}[3]{Phys.\ Lett.\ {\bf B#1} (19#2) #3}
\newcommand{\NPB}[3]{Nucl.\ Phys.\ {\bf B#1} (19#2) #3}
\newcommand{\PRD}[3]{Phys.\ Rev.\ {\bf D#1} (19#2) #3}
\newcommand{\ZPC}[3]{Zeit.\ Phys.\ {\bf C#1} (19#2) #3}
\newcommand{\SJNP}[3]{Sov.\ J.\ Nucl.\ Phys.\ {\bf #1} (19#2) #3}
\newcommand{\JETPL}[3]{JETP Lett.\ {\bf #1} (19#2) #3}

\bibitem{Georgi}                                                     
H.~Georgi. Weak Interactions and Modern Particle Theory.
Benjamin/Cummings (1984).

\bibitem{HQET}                                                       
E.~Eichten, B.~Hill. \PLB{234}{90}{511}; {\bf B243} (1990) 427;
B.~Grinstein. \NPB{339}{90}{253};
H.~Georgi. \PLB{240}{90}{447};
N.~Isgur, M.~B.~Wise. \PLB{232}{89}{113}; {\bf B237} (1990) 527.

\bibitem{HQETrev}                                                    
H.~Georgi. Proc.\ Theoretical Advanced Study Institute, Boulder, Colorado,
ed.\ R.~K.~Ellis, C.~T.~Hill, J.~D.~Lykken, World Scientific (1992), p.~598;
B.~Grinstein. Proc.\ Workshop on High Energy Phenomenology, Mexico City,
ed.\ R.~Huerta, M.~A.~P\'erez, World Scientific (1992);
Annual Reviews of Nuclear and Particle Science, v.~{\bf 42} (1992) 101;
A.~G.~Grozin. Preprint BudkerINP 92-97 (1992);
T.~Mannel. Chinese J.\ of Phys. {\bf 31} (1993) 1;
M.~Neubert. Preprint SLAC-PUB-6263 (1993).

\bibitem{HHCT}                                                       
M.~B.~Wise. \PRD{45}{92}{R2188};
G.~Burdman, J.~F.~Donoghue. \PLB{280}{92}{287};
T.-M.~Yan, H.-Y.~Cheng, C.-Y.~Cheung, G.-L.~Lin, Y.~C.~Lin, H.-L.~Yu.
\PRD{46}{92}{1148}.

\bibitem{Cho}                                                        
P.~Cho. \PLB{285}{92}{145}; \NPB{396}{93}{183}.

\bibitem{ExMes}                                                      
A.~F.~Falk, M.~Luke. \PLB{292}{92}{119};
U.~Kilian, J.~G.~K\"orner, D.~Pirjol. \PLB{288}{92}{360}.

\bibitem{Rad}                                                        
P.~Cho, H.~Georgi. \PLB{296}{92}{408};
J.~F.~Amundson, C.~G.~Boyd, E.~Jenkins, M.~Luke, A.~V.~Manohar,
J.~L.~Rosner, M.~J.~Savage, M.~B.~Wise. \PLB{296}{92}{415};
H.-Y.~Cheng, C.-Y.~Cheung, G.-L.~Lin, Y.~C.~Lin, T.-M.~Yan, H.-L.~Yu.
\PRD{47}{93}{1030}.

\bibitem{Loop}                                                       
B.~Grinstein, E.~Jenkins, A.~Manohar, M.~Savage, M.~Wise.
\NPB{380}{92}{369};
E.~Jenkins, M.~J.~Savage. \PLB{281}{92}{331};
J.~L.~Goity. \PRD{46}{92}{3929}.

\bibitem{ACCMOR}                                                     
S.~Barlag {\it et al.} (ACCMOR collaboration). \PLB{278}{92}{480}.

\bibitem{CLEO}                                                       
F.~Butler {\it et al.} (CLEO collaboration).
Preprint CLNS-92-1143 (1992).

\bibitem{SVZ}                                                        
M.~A.~Shifman, A.~I.~Vainshtein, V.~I.~Zakharov.
\NPB{147}{79}{385, 448, 519}.

\bibitem{Shuryak}                                                    
E.~V.~Shuryak. \NPB{198}{82}{83}.

\bibitem{BrGr}                                                       
D.~J.~Broadhurst, A.~G.~Grozin. \PLB{274}{92}{421};
E.~Bagan, P.~Ball, V.~M.~Braun, H.~G.~Dosh. \PLB{278}{92}{457}.

\bibitem{GrYak}                                                      
A.~G.~Grozin, O.~I.~Yakovlev. \PLB{285}{92}{254}.

\bibitem{IoSm}                                                       
B.~L.~Ioffe, A.~V.~Smilga. \JETPL{37}{83}{250}; \NPB{232}{84}{109};
I.~I.~Balitsky, A.~V.~Yung. \PLB{129}{83}{328}.

\bibitem{BelKog}                                                     
V.~M.~Belyaev, Ya.~I.~Kogan. \JETPL{37}{83}{611}; \PLB{136}{84}{273};
V.~M.~Belyaev, B.~Yu.~Blok, Ya.~I.~Kogan. \SJNP{41}{85}{439}.

\bibitem{ElKog}                                                      
V.~L.~Eletsky, Ya.~I.~Kogan. \ZPC{28}{85}{155}.

\bibitem{Use}                                                        
V.~A.~Novikov, M.~A.~Shifman, A.~I.~Vainshtein, M.~B.~Voloshin,
V.~I.~Zakharov. \NPB{237}{84}{525};
V.~L.~Chernyak, A.~R.~Zhitnitsky, I.~R.~Zhitnitsky. \SJNP{38}{83}{775}.

\end{thebibliography}
\end{document}